\documentclass[aps,jcp,twocolumn,preprintnumbers,amsmath,amssymb]{revtex4}

\usepackage{graphicx}
\usepackage{dcolumn}
\usepackage{bm}
\usepackage{footmisc}

\begin{document}

\preprint{}

\title{The quantum defect: the true measure of time-dependent 
density-functional results for atoms}

\author{Meta {\surname van Faassen}}

\author{Kieron Burke}%
\affiliation{%
Department of Chemistry and Chemical Biology, Rutgers University\\
610 Taylor Rd., Piscataway, NJ 08854-8087, USA
}%

\date{\today}

\begin{abstract}
Quantum defect theory is applied to (time-dependent) density-functional
calculations of Rydberg series for closed shell atoms: He, Be, and Ne.
The performance and behavior of such calculations is much better
quantified and understood in terms of the quantum defect, rather
than transition energies.
\end{abstract}

\maketitle

\section{\label{sec:introduction}Introduction}
Time-dependent density functional theory (TDDFT)~\cite{RG84} has
enjoyed a recent surge in popularity for calculating excited-state
energies of atoms, molecules, clusters, and solids~\cite{FR05,BWG05}.
TDDFT has features similar to ground-state density functional theory
(DFT): It produces useful accuracy at a fraction of the computational
cost of ab initio methods~\cite{FR05}, but reliability depends on the
approximate functionals used~\cite{FB04}.  As more and more
practictioners in many subfields of computational science use TDDFT,
there are ever increasing numbers of implementations.  Calculations on
atoms and sequences of atoms are often used to benchmark new
implementations, or to show how well TDDFT works in the simplest
cases.  These are particularly useful, as much highly accurate data,
both from experiment and accurate wave function calculations, are
available for these systems.

A well-known difficulty that hampered even the earliest calculations
of excitations in atoms with TDDFT~\cite{PGG96} is the incorrect
asymptotic behavior of the ground-state potentials of common density
functional approximations.  It has long been known that the exact KS
potential for a closed shell atom decays as $-1/r$ at large $r$, where
$r$ is the distance from the nucleus~\cite{LPS84,AP84}.  Typical
approximations for the ground-state, such as the local density
approximation (LDA), gradient corrected functionals (GGAs), and hybrid
functionals (see for example Ref.~\cite{M02} and references therein),
have potentials that decay too rapidly with $r$.  Only those with the
correct asymptotic behavior support the Rydberg series of transitions,
an infinite number of transitions that merge with the continuum at the
ionization threshold.  Orbital-dependent functionals capture this
behavior naturally, as does the Van Leeuwen-Baerends potential
approximation (LB94)~\cite{LB94}, which was designed to have an
asymptotically correct behavior.  Several recent methods have been
suggested for correcting the standard functionals to produce the
long-ranged tail~\cite{THb98,CS00,WY03}.  In any event, our present
work applies only to long-ranged potentials.

We argue here that the way in which results have been calculated and
reported for atoms is far from optimal.  We show that long lists of
transition frequencies for Rydberg series converging to the ionization
threshold are {\em not} the best way to report such calculations.
Instead, the well-developed theory of the quantum defect~\cite{F98},
used for decades in atomic physics, is ideal for this purpose.  We
show that, for each Rydberg series, i.e., for each value of angular
momentum $l$, two or three numbers completely characterize all the
information in the infinite series.  Furthermore, the quantum defect
is a much more demanding test of excitation energies, and methods that
appear to have only small energetic errors can yield quite poor
quantum defect behavior.  Also, shifts in orbital energies, such as
the missing correlation contribution in the exact exchange HOMO
energy, have no effect on the quantum defect, so that the quality of a
potential can be assessed without being influenced by such errors.  We
also find~\cite{ARU98}, that the quantum defect of the exact
ground-state Kohn-Sham (KS) potential, is sometimes, but not always, a good
starting point for approximations to the true quantum defect.
\begin{figure}
\includegraphics[width=3.3in]{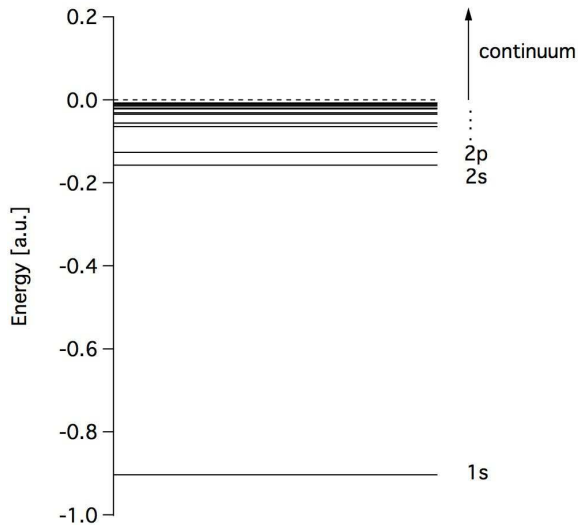}
\caption{\label{fig:helevels} Energy level diagram for the helium atom.}
\end{figure}

\section{\label{sec:theory}Theory}
In Fig.~\ref{fig:helevels} we show the orbital energy level diagram of
the helium atom.  The zero is set at the onset of the continuum, marked
with a dotted
line. For closed shell atoms and
for any spherical one-electron potential that decays as $-1/r$ at
large distances, the bound-state transitions form a Rydberg series
with frequencies:
\begin{equation}
\omega_{nl}=I-\frac{1}{2(n-\mu_{nl})^2}
\label{eq:qd}
\end{equation}
where $I$ is the 
ionization potential, and
$\mu_{nl}$ is called the quantum defect.
We use atomic units ($e^2=\hbar=m_e=1$) throughout.
The value of the quantum defect is that, for real atoms,   quantum defects
depend only weakly on the principle quantum number $n$ for large $n$
and converge to a finite value in the limit $n\rightarrow\infty$.
In fact, according to Seaton's theorem~\cite{S58}, the quantum 
defect is a smooth function of energy as $E\to 0$, and merges
continuously with the phase-shift (relative to pure Coulomb scattering)
divided by $\pi$.
In Table \ref{tab:helium}, we report extremely accurate results from
wavefunction calculations for the helium atom. We show singlet and
triplet values that have been obtained by Drake~\cite{D96}. We also
give results from the exact ground-state KS potential, as found
by Umrigar and Gonze~\cite{UG94}. We say more on how we obtained the
KS values in section~\ref{sec:comput}.  On the left are the
transition frequencies, while on the right are the corresponding
quantum defects.  Note how small the differences between transitions
become as one climbs up the ladder, and yet the quantum defect remains
finite and converges to a definite value.
\begingroup
\squeezetable
\begin{table*}
\caption{\label{tab:helium} Transition energies for He atom [a.u.]. 
The ionization energies are 0.9037244 a.u. for the singlet and triplet case 
and 0.90372 a.u. for the KS potential.}
\begin{ruledtabular}
\begin{tabular}{cdddd|ddd}
\multicolumn{1}{c}{} &
\multicolumn{4}{c|}{$\Delta E$} &
\multicolumn{3}{c}{Quantum defect} \\
  \multicolumn{1}{c}{Transition} & \multicolumn{1}{c}{Singlet \footnotemark[1]} & 
\multicolumn{1}{c}{Triplet \footnotemark[1]} & \multicolumn{1}{c}{KS\footnotemark[2]} 
& \multicolumn{1}{c|}{KS fit} & \multicolumn{1}{c}{ Singlet} 
& \multicolumn{1}{c}{Triplet} & \multicolumn{1}{c}{KS}\\\hline  
  $1s\rightarrow 2s$ & 0.7577503  & 0.7284949 & 0.74599 & 0.74596 &
  0.1492525 & 0.3107982 & 0.21957\\
  $1s\rightarrow 3s$ & 0.8424524  & 0.8350353 & 0.83917 & 0.83920 &
  0.1433699 & 0.3020047 & 0.21689\\
  $1s\rightarrow 4s$ & 0.8701376  & 0.8672122 & 0.86883 & 0.86882 &
  0.1416535 & 0.2994464 & 0.21492\\
  $1s\rightarrow 5s$ & 0.8825475  & 0.8883469 & 0.88189 & 0.88189 &
  0.1409171 & 0.2983583 & 0.21459\\
  $1s\rightarrow 6s$ & 0.8891613  & 0.8925944 & 0.88879 & 0.88879 &
  0.1405330 & 0.2977954 & 0.21441\\
  $1s\rightarrow 7s$ & 0.8930986 & 0.8925945 & 0.89287 & 0.89287 &
  0.1403072 & 0.2974666 & 0.21429
\end{tabular}
\end{ruledtabular}
\footnotetext[1]{Accurate non-relativistic calculations from Ref.~\cite{D96}.}
\footnotetext[2]{The differences between the KS eigenvalues obtained 
with the exact potential from Ref.~\cite{UG94}.}
\end{table*}
\endgroup

\begin{figure}
\includegraphics[width=3.3in]{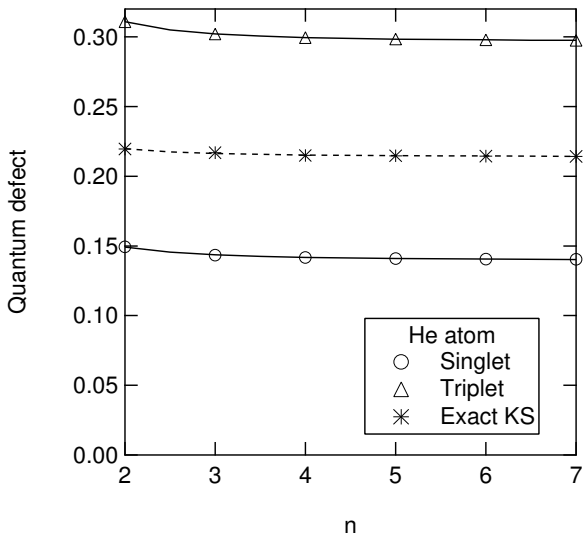}
\caption{\label{fig:stks} The exact $s$ KS quantum defect and the exact 
singlet and triplet quantum defects~\cite{D96} of He and their parabolic fits.}
\end{figure}
In Fig.~\ref{fig:stks} we show the exact $s$ KS quantum defect
and the singlet and triplet quantum defects corresponding to accurate
wave function results~\cite{D96} for helium as symbols.  This is the way the
defect was plotted in Ref.~\cite{ARU98} for the case of the Ne atom.
\begin{figure}
\includegraphics[width=3.3in]{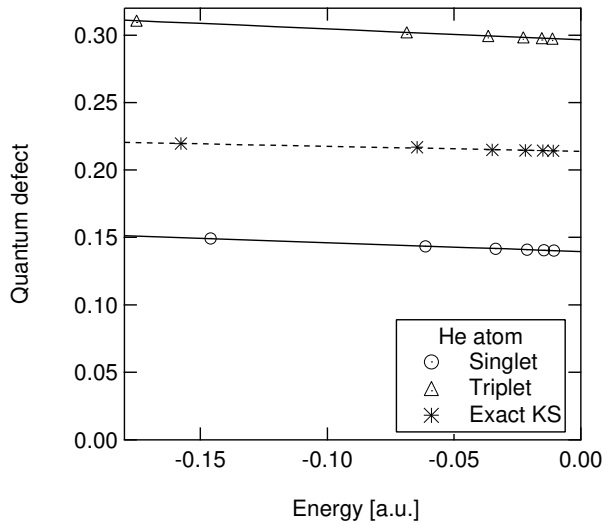}
\caption{\label{fig:exactvse} Same as Figure~\ref{fig:stks} plotted against energy.}
\end{figure}
However, it is more appropriate still to plot the defect as a function
of energy, as shown in Fig.~\ref{fig:exactvse}.  This clearly illustrates that the
quantum defect is a smooth function of energy, and will be well
approximated a polynomial of some low order $p$:
\begin{equation}\label{eq:fit}
\mu^{(p)}(E) =\sum_{i=0}^p \mu_i E^i,\quad E=\omega-I.
\end{equation}
We choose to optimize the fit over the entire range of
exitation energies, not just about $E=0$ (the $\mu_i$ are
simply related to the $a$, $b$, and $c$ coefficients in
Ref.~\cite{ARU98}). For example the data in Fig.~\ref{fig:exactvse} can be accurately described by
a straight line.
\begin{table}
\caption{\label{tab:heab} Fit coefficients $\mu_0$, and $\mu_1$ for the $s$ quantum
 defect of He. By `Max. AE' we mean the maximum absolute error as explained in the text.}
\begin{ruledtabular}
\begin{tabular}{cddd}
  \multicolumn{1}{c}{} & \multicolumn{1}{c}{Singlet\footnotemark[1]} & 
\multicolumn{1}{c}{KS\footnotemark[2]}& \multicolumn{1}{c}{Triplet\footnotemark[1]}\\\hline  
 $\mu_0$ & 0.1396 & 0.2139 & 0.2965\\
 $\mu_1$ & -0.0655 & -0.0370 & -0.0811\\ 
Max. AE & 0.0002 & 0.0006 & 0.0001
\footnotetext[1]{Obtained from non-relativistic calculations of the orbital 
energies from Ref.~\cite{D96}.}
\footnotetext[2]{KS values obtained with the exact potential from Ref.~\cite{UG94}.}
\end{tabular}
\end{ruledtabular}
\end{table}
We report these coefficients in Table~\ref{tab:heab} and give the
quantum defects obtained from these coefficients as continuous lines
in Figs.~\ref{fig:stks} and~\ref{fig:exactvse}. We also added the
transition energies for He obtained from the KS fit to
Table~\ref{tab:helium}. The maximun error in the fit is 3/100 mH so it
is essentially exact for all purposes of this paper.  We shall see
that in other cases just two or even three coefficients are not enough
to describe the data accurately. In those cases we will need to decide
when to stop adding more coefficients to fit the data, since we would
like to describe the data with as few coefficients as possible.
Therefore we will look at the maximum absolute error in the quantum
defect. By which we mean that we recalculate the quantum defects from
the coefficients and look at the absolute differences between the
fitted and original values, and then look at the largest difference.
We will stop adding more coefficients once the maximum absolute error
is smaller than 0.001, or, in case this value is not reached, whenever
the absolute errors do not change much when adding more coefficients.
We give the value for the error also in Table~\ref{tab:heab}.  Any
approximate ground-state KS potential suggested for use in TDDFT
should have its coefficients compared with the KS numbers in the
table, while any approximate xc-kernel should have its coefficients
compared with the singlet and triplet case.

\section{\label{sec:groundstate} Ground-state Kohn-Sham potentials}

All linear response TDDFT calculations of excitations begin from the occupied to
unoccupied transitions of the {\em ground-state} KS potential. In this
section, we analyse some of the most popular approximations using
quantum defect theory.

\subsection{\label{sec:comput} Computational details}
All ground-state DFT results shown here are calculated with a modified
OEP (optimized potential model) program~\cite{Engel,ED99,TS76}. This
program is basis set independent, works with a radial grid, and both
the energies and the potentials are optimized in a self-consistent
way. The exact-exchange (x-only) OEP is already included in this
program. We also did calculations with the LB94~\cite{LB94} potential
by Van Leeuwen and Baerends, which was not available in the program.
We implemented this functional by adding the LB94 correction to the
LDA xc-potential and let the program optimize this potential in a self
consistent manner. The program has the ability to read in the accurate
potentials by Umrigar {\em et al.}  (He~\cite{UG94}, Be~\cite{UG93},
and Ne~\cite{UG93}). These accurate potentials were only known by us
up to a particular radius. In order to calculate energies for higher
$n$ values we needed to increase this radius. We did this by adding a
$-1/r$ tail to $\upsilon_{\rm xc}$ and a $Z/r$ tail to $\upsilon_{rm
  H}$ and we checked that the transition was smooth. We also made sure
our values are converged with the number of gridpoints. When we used
the accurate potentials, we did not allow the program to self
consistently change the potential.

The maximum $n$ value for which we could still do very accurate calculations
is  $n=7$ for He and $n=9$ for Be and Ne.

\subsection{\label{sec:exactpots}Exact results}
As we have seen the  numbers, $\mu_i$, contain all the information
needed to characterize a given Rydberg series, and make tables of the
actual transition frequencies redundant.  
\begin{table}
\caption{ \label{tab:umrigarabc} The $\mu_i$ from  the exact ground-state KS 
potential for He, Be, and Ne.}
\begin{ruledtabular}
\begin{tabular}{cddd}
  \multicolumn{1}{c}{} & \multicolumn{1}{c}{He\footnotemark[1]} &  
\multicolumn{1}{c}{Be\footnotemark[2]}  &  \multicolumn{1}{c}{Ne\footnotemark[2]}  \\\hline  
   \multicolumn{4}{c}{$s$}\\\hline
   I     & 0.9037 & 0.3426 &  0.7945\\  
 $\mu_0$ & 0.2139 & 0.7164 & 1.3125\\
  $\mu_1$ & -0.0370 & -0.2185 & -0.1807\\
   Max. AE & 0.0006 & 0.0002 & 0.0007 \\\hline
    \multicolumn{4}{c}{$p$}\\\hline 
     $\mu_0$  & 0.0164 & 0.3587 & 0.8304\\
  $\mu_1$  & 0.0289 & -0.3377 & -0.3500\\
   $\mu_2$ &  & 0.6112 & \\
   Max. AE & 0.0009 & 0.0003 & 0.001
 \end{tabular}
\end{ruledtabular}
\footnotetext[1]{Ref.~\cite{UG94}}
\footnotetext[2]{Ref.~\cite{UG93}}
\end{table}
In Table~\ref{tab:umrigarabc}, we report the coefficients for $s$ and
$p$ KS quantum defects for the He, Be and Ne atoms obtained with accurate xc
potentials. We use the transitions up to $n=9$ for the fit in case of
Be and Ne, and up to $n=7$ in case of He. 

When we compare the $s$ ($l=0$) values with $p$ ($l=1$), the
asymptotic KS quantum defect is smaller for $p$ in all cases.  This
reflects the lesser importance of the inner part of the KS potential
relative to the angular momentum barrier as $l$ grows.  However, the
curvature of the $s$ and $p$ quantum defect is similar.

\subsection{\label{sec:kspots} Approximations}
In this section, we demonstrate our methodology by testing two common
approximations for the ground-state KS potential. These are exact
exchange OEP~\cite{TS76} and LB94~\cite{LB94}. Exact exchange
calculations are more demanding than traditional DFT calculations, but
are becoming popular because of the high quality of the
potential~\cite{G99, IHB99}. On the other hand, LB94 provides an
asymptotically correct potential at little extra cost beyond
traditional DFT~\cite{CS00,GGGB01,WAY03}.
\begin{figure}
\includegraphics[width=3.3in]{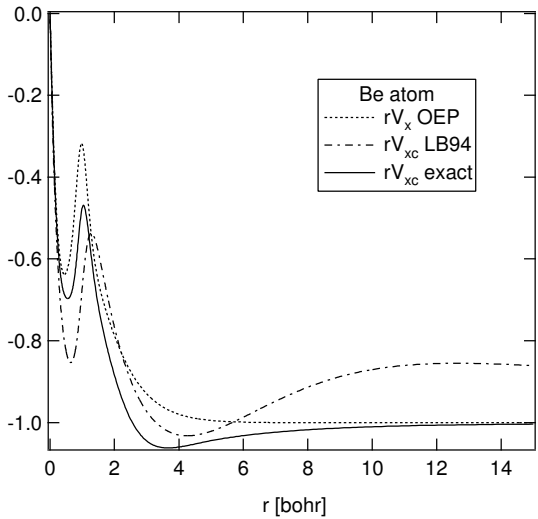}
\caption{\label{fig:bekspots} The exact, LB94, and x-only potentials for 
beryllium together with $-1/r$.}
\end{figure}
Fig.~\ref{fig:bekspots} shows both these potentials for the Be atom
together with the exact potential~\cite{UG93}. Our figure for the
exact potential is different from the one in Ref.~\cite{LB94}, since
in that reference they use a different accurate density to derive the
potential.
\begin{table*}
\caption{\label{tab:kspotsabc} The $\mu_i$ from different ground-state
 potentials for He, Be, and Ne. The ionization energies are in [a.u.] and are 
 not included in the maximum error. }
\begin{ruledtabular}
\begin{tabular}{ccddd|ddd}
  \multicolumn{2}{c}{} & \multicolumn{3}{c|}{$s$}  & \multicolumn{3}{c}{$p$} \\
  \multicolumn{2}{c}{} & \multicolumn{1}{c}{OEP} & \multicolumn{1}{c}{LB94}& 
   \multicolumn{1}{c|}{exact} & \multicolumn{1}{c}{OEP} & \multicolumn{1}{c}{LB94} & 
   \multicolumn{1}{c}{exact} \\\hline  
  He   & $I$ & 0.9182 & 0.8513 & 0.9037 \\
       & $\mu_0$ & 0.2170 & -0.3607 & 0.2139 & 0.0238 & -0.6015 & 0.0164\\
       & $\mu_1$ & -0.0401 & -2.7408 & -0.0370 & 0.0319 &  -2.2013 & 0.0289\\
       & $\mu_2$ & & -3.1852 && & -1.5723 &\\
       & Max. AE & 0.0000 & 0.003 & 0.0006 & 0.0000 & 0.002 & 0.0009\\ \hline
  Be   & $I$ & 0.3093 & 0.3205 & 0.3426 \\
       & $\mu_0$ & 0.6623 & 0.3497 & 0.7164 & 0.2786 & 0.0115 & 0.3287\\
       & $\mu_1$ & -0.1557 & -0.7203 & -0.2185 & -0.1070 & 0.4836 & -0.3377\\
       & $\mu_2$ & & 31.5313 &  & 0.8453 & 66.0517 & 0.6112\\
       & $\mu_3$ & & 73.1572 &  &  & 268.6094 & \\
       & Max. AE & 0.0002 &  0.001 & 0.0002 & 0.0002 &  0.002 & 0.0003\\ \hline
  Ne   & $I$ & 0.8508 & 0.7821 & 0.7945 \\ 
       & $\mu_0$ & 1.3433 & 0.8174 & 1.3125 & 0.8631 & 0.3196 &  0.8340\\
       & $\mu_1$ & -0.2285 & -3.2880 & -0.1807& -0.3972 & -2.8338 & -0.3500\\
       & $\mu_2$ && -11.0968 & && -1.3271 & \\
       & $\mu_3$ &  & -39.2359 &  &  &  & \\
       & Max. AE & 0.0004 & 0.004 & 0.0007 & 0.0008 & 0.003 & 0.001
\end{tabular}
\end{ruledtabular}
\end{table*}
We give our values for the fit parameters  in
Table~\ref{tab:kspotsabc}.
 \begin{figure}
\includegraphics[width=3.3in]{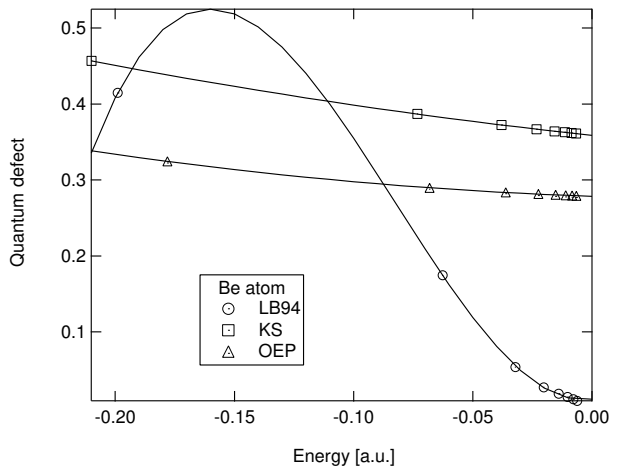}
\caption{\label{fig:lb94fit} The Be $p$ quantum defect of LB94, OEP, 
and KS, and their best fits.}
\end{figure}
In Fig.~\ref{fig:lb94fit} we show the $p$ Be quantum defect obtained
with LB94, OEP, and KS and we show the fit as continuous lines.
Fig.~\ref{fig:lb94fit} immediately shows the high quality of the OEP
potential. The quantum defect curve is almost identical to the exact
one, being offset by about 0.1 (see Table~\ref{tab:kspotsabc}). On the
other hand the quantum defect of LB94 is poor, and this is true for
all cases studied.  From Fig.~\ref{fig:bekspots} we can see that the
OEP is much closer to the true potential than LB94 and also approaches
$-1/r$ faster. This shows that just having a potential that is
asymptotically correct is not enough to get a good quantum defect.

Another thing we have not mentioned so far is that a potential that
gives a wrong ionization potential does not necessarily give a bad
quantum defect. The origin of this is that the quantum defect is
obtained from $\omega_{nl}-I$ and can take a shift of energy levels
into account.  Consider the exact exchange results for the atoms.
Typically, these quantum defects are accurate to 0.1.  Thus, using
the exact ionization potential with the exchange quantum defects,
yields highly accurate transition frequencies, i.e., the most
significant error in OEP excitations is due to the missing
correlation contribution to the position of the HOMO.  On the other
hand, we see that LB94, while asymptotically correct and sometimes
having a highly accurate ionization potential, has much less accurate
quantum defects.

\section{\label{sec:results}  TDDFT Results}
\begin{table*}
\caption{\label{tab:beryllium}Transition energies for the beryllium atom.}
\begin{ruledtabular}
\begin{tabular}{cccccccc}
  \multicolumn{4}{c}{}& \multicolumn{1}{c}{Truncated-}&\multicolumn{1}{c}{Truncated-}
  &\multicolumn{2}{c}{}\\
  Transition & Expt.\footnotemark[1] & KS\footnotemark[2] &  
  ALDA\footnotemark[3] & ALDA\footnotemark[4] &  Hybrid\footnotemark[5] & 
  WY\footnotemark[6] & AC-LDA\footnotemark[7]  \\\hline 
  $2s\rightarrow 3s$ & 0.24913 & 0.24437 & 0.2495 & 0.2515 & 0.2510 & 0.2512 & 0.2239\\
  $2s\rightarrow 4s$ & 0.29728 & 0.29586 & 0.2977 & 0.2984 & 0.2985 & 0.3033 & 0.2680\\
  $2s\rightarrow 5s$ & 0.31586 & 0.31526 & 0.3160 & 0.3164 & 0.3165 & &  \\
  $2s\rightarrow 6s$ & 0.32496 & 0.32466 &        & 0.3252 & 0.3254 & & \\
  \\
  $2s\rightarrow 2p$ & 0.19394 & 0.13273 & 0.1868 & 0.1889 & 0.1427 & 0.1861 & 0.1781\\
  $2s\rightarrow 3p$ & 0.27423 & 0.26937 & 0.2710 & 0.2714 & 0.2736 & 0.2749 & 0.2457\\
  $2s\rightarrow 4p$ & 0.30543 & 0.30461 & 0.3048 & 0.3049 & 0.3059 & 0.3107 & 0.2754\\
  $2s\rightarrow 5p$ & 0.31949 & 0.31931 &        & 0.3194 & 0.3199 &        & \\
  $2s\rightarrow 6p$ & 0.32690 & 0.32686 &        & 0.3269 & 0.3272 &        & \\
  \\
  $s$ MAE [mH] & & & 0.312 & 1.069 & 0.943 & 4.046 & 27.25 \\
  $p$ MAE [mH] & &  & 3.669 & 1.700 & 10.61 & 4.592 & 24.81 \\
  Total MAE [mH] & & & 1.991 & 1.420 & 6.313 & 4.374 & 25.79
\end{tabular}
\end{ruledtabular}
\footnotetext[1]{Experimental values from NIST~\cite{NIST}}
\footnotetext[2]{The differences between the KS eigenvalues obtained with 
the exact potential from Ref.~\cite{UG93}\label{fn:KS}}
\footnotetext[3]{ALDA calculation including all bound and unbound states from Ref.~\cite{GKSG98}}
\footnotetext[4]{ALDA calculation including 34 unbound states from Ref.~\cite{PGB00}}
\footnotetext[5]{Hybrid calculation including 34 unbound states for He and 
38 unbound states for Be from Ref.~\cite{BPG00}}
\footnotetext[6]{ALDA calculation with WY ground-state potential from Ref.~\cite{WCY05}}
\footnotetext[7]{Asymptotically corrected ALDA results from Ref.~\cite{WCY05}}
\end{table*}
In the previous sections we saw that the KS quantum defects are
typically lying in between the exact singlet and triplet quantum
defects~\cite{SUG98}. In order to calculate these singlet and triplet quantum
defects within DFT the usual method of choice is TDDFT (within the
linear response regime).  Apart from a ground-state potential one
needs to choose an xc-kernel as well. In this section we obtain
quantum defects from excitation energies obtained with different
xc-kernels and ground-state potentials.  We focus mainly on the
Be atom, but we also give expansion coefficients for He and Ne quantum
defects.

\begin{table*}
\caption{\label{tab:BetddftabcS} The $\mu_i$ from different singlet TDDFT values 
for Be.} 
\begin{ruledtabular}
\begin{tabular}{cddddddd}
  \multicolumn{1}{c}{} & \multicolumn{1}{c}{Ref.\footnotemark[1]}
& \multicolumn{1}{c}{KS\footnotemark[2]}
 & \multicolumn{1}{c}{ALDA\footnotemark[3]}
& \multicolumn{1}{c}{Truncated-ALDA\footnotemark[4]} 
& \multicolumn{1}{c}{Truncated-Hybrid\footnotemark[5]}
&  \multicolumn{1}{c}{WY/ALDA\footnotemark[6]}
& \multicolumn{1}{c}{AC-LDA\footnotemark[7]}  
\\ \hline  
  \multicolumn{1}{c}{} & \multicolumn{7}{c}{$s$} \\ \hline
  $I$ & 0.3426 & 0.3426 & 0.3426 & 0.3426 & 0.3426  & 0.3493 & 0.3111 \\  
  $\mu_0$ & 0.6752 & 0.7164 & 0.6536 & 0.6182 & 0.6010  &  0.6684 & 0.5997\\
  $\mu_1$ & -0.0133 & -0.2185  & -0.2986 & -0.4294 &  -0.6947  &  -0.7540 & \\
  $\mu_2$ & 1.2345 &  &  &  &  & \\
 Max. AE & 0.0003 & 0.0002 & 0.004 &  0.002 & 0.004 & 0.0000 & 0.006  \\ \hline
  \multicolumn{1}{c}{} & \multicolumn{7}{c}{$p$} \\ \hline  
   $\mu_0$ & 0.3745 & 0.3287 & 0.3327 & 0.3427 & 0.3087  & 0.3484 & 0.2413 \\
  $\mu_1$ & 1.1612 & -0.3377 & -1.3172 & -0.9634 & 0.2505  & -1.9702  & -1.1215 \\
  $\mu_2$ & 1.5678 & 0.6112 & -13.5673 & -12.4534 & 3.9961 & -15.7792  & -18.6196  \\
  $\mu_3$ & 21.4587 &  &  &  &  &  \\
  Max. AE & 0.0002 & 0.0003 & 0.0000 & 0.003 & 0.004  & 0.0000 &  0.0000
\end{tabular}
\end{ruledtabular}
\footnotetext[1]{Experimental values from NIST~\cite{NIST}}
\footnotetext[2]{The differences between the KS eigenvalues obtained with 
the exact potential from Ref.~\cite{UG93}}
\footnotetext[3]{ALDA calculation including all bound and unbound states from Ref.~\cite{GKSG98}}
\footnotetext[4]{ALDA calculation including 34 unbound states from Ref.~\cite{PGB00}}
\footnotetext[5]{Hybrid calculation including 38 unbound states from Ref.~\cite{BPG00}}
\footnotetext[6]{ALDA calculation with WY ground-state potential from Ref.~\cite{WCY05}}
\footnotetext[7]{Asymptotically corrected ALDA results from Ref.~\cite{WCY05}}
\end{table*}
\subsection{\label{sec:perftddft}Performance of TDDFT}
We concentrate on the quantum defect obtained from the ALDA kernel
applied to the exact ground-state KS potential. In
Table~\ref{tab:beryllium}, we show $s\rightarrow s$ and $s\rightarrow
p$ excitation energies for the Be atom. The ALDA results in column
three are obtained by Van Gisbergen {\em et.  al.}~\cite{GKSG98}. For
their ground-state calculations, they used the accurate potential by
Umrigar and Gonze~\cite{UG93} and for the xc-kernel they used the
ALDA. The calculations were done close to the basis set limit and with
high numerical integration accuracy. As can be seen from the table the
excitation energies for the $s\rightarrow s$ transitions are very
close to the experimental values with a mean average error (MAE) of
only 0.3 mH. The $s\rightarrow p$ excitation energies are a bit less
accurate with an MAE of 3.7 mH, but this is still an accurate result.
The fit coefficients for the quantum defects of these calculations are
reported in Table~\ref{tab:BetddftabcS}. For the fit of the TDDFT
results we took a less strict constraint to determine when to stop
adding more coefficients. We took 0.01 instead of 0.001, reflecting
the greater error in this data.

\begin{figure}
\includegraphics[width=3.3in]{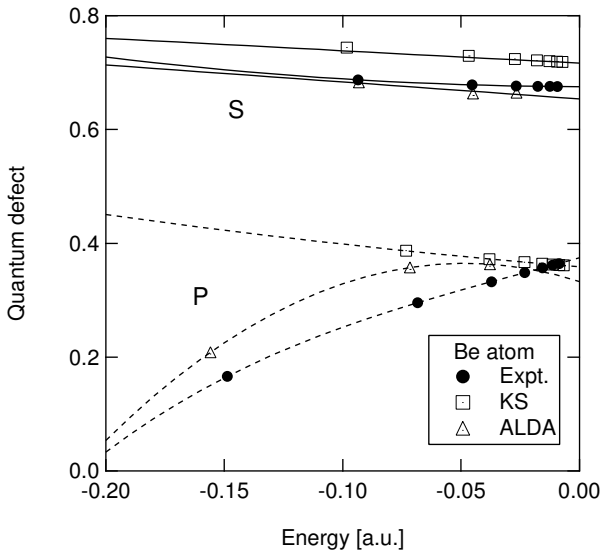}
\caption{\label{fig:aldaplot} The $s$ and $p$ experimental, KS, and ALDA 
  singlet quantum defects for Be. The continuous lines are the
  $\mu^{(p)}(E)$ fits.}
\end{figure}
In Fig.~\ref{fig:aldaplot} we show the $s$ and $p$ quantum defects
corresponding to these values and we compare them with the bare KS and
experimental results. For the $s$ quantum defect, the experimental
curve is essentially a straight line, with a small negative slope. The
$p$ quantum defect, on the other hand, is much more curved, with a
large positive slope, due to the much lower $2s\rightarrow 2p$
transition. The exact KS potential has been touted as a good
approximation to the experimental results~\cite{SUG98}. This is
clearly true for the $s$ curves but not so for the $p$ quantum defect.
In that case the KS quantum defect, while being very close to the
expermental value as $n\rightarrow\infty$ has the wrong behavior as a
function of $E$.

In both the $s$ and $p$ case it is clear that doing a full TDDFT
calculation considerably improves upon the bare KS results. The ALDA
does slightly overcorrect the $s$ quantum defect, underestimating the
value at $E=0$.  In case of the $p$ quantum defect the ALDA tends to
correct for the opposite slope of the KS values compared to the
experimental values, but the correction is not complete, leading to a
curved line.  We also see that the $s$ quantum defects as obtained
with the ALDA are much better than the $p$ quantum defects, even
though the MAE is only a few mH in the last case.

\subsection{\label{sec:truncation}Truncating Casida's equation}
Another set of ALDA calculations were performed by Petersilka, Gross,
and Burke~\cite{PGB00}. The difference between their calculation and
that of Van Gisbergen is that they truncate the summation over states
in the response function, including only poles of bound states and
neglecting continuum contributions. They included the lowest 34 bound
states of $s$ and $p$ symmetry in their calculations. Just like in
case of the full ALDA calculations the ground-state was determined
with the potential of Umrigar and Gonze~\cite{UG93}. We show these
results in column four of Table~\ref{tab:beryllium}. Again the results
are close to the experimental values with an MAE of 1.1 mH for the
$s\rightarrow s$ transitions and an MAE of 1.7 mH for the
$s\rightarrow p$ transitions.
\begin{figure}
\includegraphics[width=3.3in]{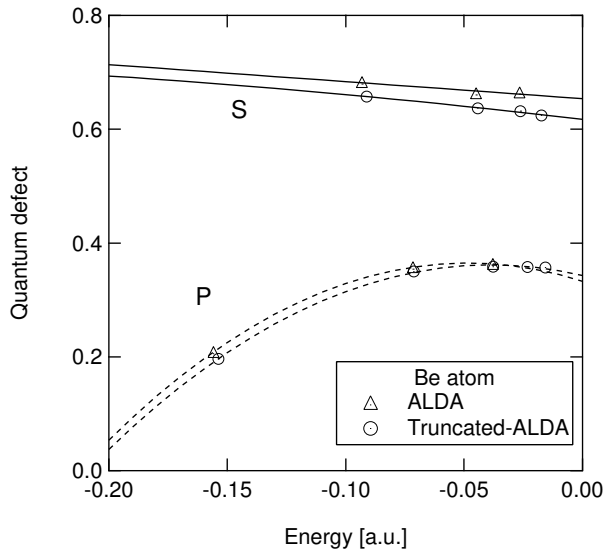}
\caption{\label{fig:truncateplot} The $s$ and $p$ ALDA and truncated-ALDA
singlet quantum defects for Be. The continuous lines are the
  $\mu^{(p)}(E)$ fits.}
\end{figure}
In Fig.~\ref{fig:truncateplot} we show the $s$ and $p$ quantum defects
corresponding to these values and we compare them with the full ALDA
results.  For the $s$ quantum defect we see that the quantum defect is
slightly below the ALDA values, and the slope is too large, leading to
a smaller asymptotic quantum defect. For the $p$ quantum defect we see
that there is not much difference between the truncated and the full
results. So the effect of the truncation in case of Be is not so
great. For He the difference is relatively larger, because the TDDFT
corrections are so small. This can be seen from the values in
Tables~\ref{tab:BetddftabcS} and~\ref{tab:HetddftabcS}, which we will
discuss in section~\ref{sec:coeffs}.

\subsection{\label{sec:hybrid}Quality of the ground-state potential}
For all TDDFT methods we have described so far, the ground-state was
calculated with the exact potential of Umrigar and Gonze~\cite{UG93}.
This eliminates errors in the ground-state so one can compare the
effects of using a different xc-kernel. But for practical calculations
such an accurate potential is not available. This motivates the
development of other accurate ground-state potentials to calculate the
Rydberg series.

Wu and Yang~\cite{WY03} obtained an accurate potential by a direct
optimization method that allows them to calculate the potential from a
given electronic density. This density is obtained from a
coupled-cluster singles and doubles (CCSD) calculation. In
Table~\ref{tab:beryllium} we show the excitation energies
corresponding to the WY potentials and the error is just a few mH.
\begin{figure}
\includegraphics[width=3.3in]{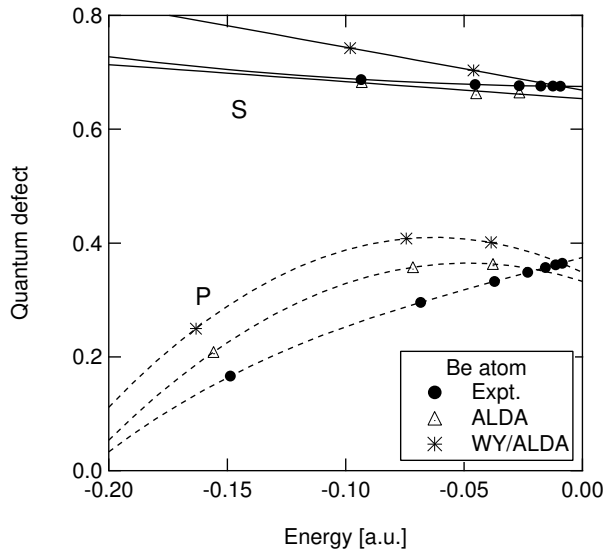}
\caption{\label{fig:wyplot} The $s$ and $p$ experimental, ALDA,  
and WY quantum defects for Be. The continuous lines are the
  $\mu^{(p)}(E)$ fits.}
\end{figure}
In Fig.~\ref{fig:wyplot} we show the quantum defects obtained with
this potentials and compare them with the ALDA and experimental
values. Only the asymptotic value of the quantum defect is accurate.
Errors due to (very small) errors in the ground-state density are
visible for all other energies, and can be comparable to the xc-kernel
itself.  The WY method is therefore still very promising, but clearly
requires a very accurate input density. Quantum defect analysis should
prove very useful for testing WY-type calculations, for example, for
comparing basis set errors with errors due to the level of calculation
used for obtaining the input density.

\subsection{\label{sec:hybrid}Testing approximate kernels}
In an attempt to improve the xc-kernel, Burke, Petersilka, and
Gross~\cite{BPG00} suggested the following form,
\begin{equation}
f_{\rm xc}^{\uparrow\uparrow}=f_{\rm x}^{\uparrow\uparrow},\quad
f_{\rm xc}^{\uparrow\downarrow}=f_{\rm xc}^{\uparrow\downarrow{\rm ALDA}}.
\end{equation}
This form is based on the fact that the parallel-spin contribution is
well described in the exact exchange case, because of a cancellation
of exchange contributions. But an exact exchange treatment misses the
significant anti-parallel correlation contribution, leading to
too large singlet/triplet splittings. Therefore it is recommended that
for the anti-parallel kernel one uses the ALDA. We show the excitation
energies obtained with this kernel in Table~\ref{tab:beryllium}. It
should be noted that just as in case of the truncated-ALDA
calculations, the number of states included in the hybrid calculation
is also limited. Namely, 34 states in case of He and 38 states in case
of Be. Therefore we shall denote the method by truncated-hybrid. The
ground-state of these calculations was again calculated with the exact
potential of Umrigar and Gonze~\cite{UG93}. Apart from the
$2s\rightarrow 2p$ transition the truncated-hybrid results are very
similar to the truncated-ALDA results. It is the large error in the
$2s\rightarrow2p$ transition that leads to the large MAE.

\begin{figure}
\includegraphics[width=3.3in]{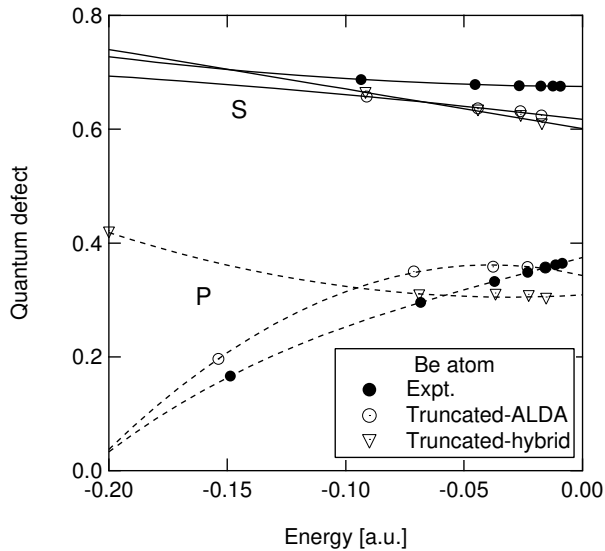}
\caption{\label{fig:hybridplot} The $s$ and $p$ experimental, truncated-ALDA, and 
  truncated-hybrid singlet quantum defects for Be. The continuous
  lines are the $\mu^{(p)}(E)$ fits.}
\end{figure}
In Fig.~\ref{fig:hybridplot} we show the $s$ and $p$ quantum defects
corresponding to the truncated-hybrid results and we compare them with
the truncated-ALDA and experimental results. From the $s$ quantum
defect plot it can again be seen that the truncated-hybrid results are
close to the truncated-ALDA values, so the kernel does not improve the
results in this case.  For the $p$ quantum defect, the results actually
get worse with the truncated-hybrid giving a slope opposite the the
experimental curve, so it shifts the KS values in the right direction
but does not correct for the wrong slope. The hybrid kernel does not
improve much upon the truncated-ALDA for Be as was also found in
Ref.~\cite{BPG00}. It is only good for two-electron systems, for which
it was derived.

\subsection{\label{sec:wy}Asymptotically corrected ground-state potentials}
Wu, Ayers, and Yang~\cite{WAY03} obtained an asymptotically corrected
LDA (AC-LDA) potential by a variational method that forces the
potential to have the correct asymptotic behavior. This is a pure DFT
treatment that can be applied to larger molecules. We see from
Table~\ref{tab:beryllium} that the AC-LDA gives a large MAE.

\begin{figure}
\includegraphics[width=3.3in]{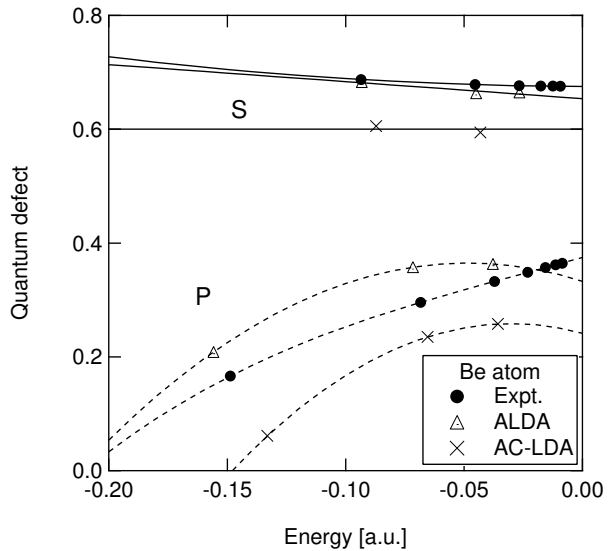}
\caption{\label{fig:acldaplot} The $s$ and $p$ experimental, ALDA, 
and AC-ALDA quantum defects for Be. The continuous lines are the
  $\mu^{(p)}(E)$ fits.}
\end{figure}
In Fig.~\ref{fig:acldaplot} we show the quantum defects obtained with
the AC-LDA potential and compare them with the KS and experimental
values.  In both cases the AC-LDA values are much lower than the ALDA,
which was evaluated on the exact ground-state KS potential.  The
AC-LDA strongly underestimates the experimental $s$ and $p$ quantum
defects.  The shape of the $p$ curve is similar to the ALDA curve.
From this figure, it is clear that there are significant errors in the
underlying KS potential.
\subsection{\label{sec:coeffs}Coefficients for He}
\begin{table*}
\caption{\label{tab:HetddftabcS} The $\mu_i$ from different singlet TDDFT values 
for He.} 
\begin{ruledtabular}
\begin{tabular}{cddddddd}
  \multicolumn{1}{c}{} & \multicolumn{1}{c}{Ref.\footnotemark[1]} 
&  \multicolumn{1}{c}{KS\footnotemark[2]}
& \multicolumn{1}{c}{ALDA\footnotemark[3]}
& \multicolumn{1}{c}{Truncated-ALDA\footnotemark[4]} 
& \multicolumn{1}{c}{Truncated-Hybrid\footnotemark[5]} 
 &  \multicolumn{1}{c}{WY/ALDA\footnotemark[6]} 
& \multicolumn{1}{c}{AC-LDA\footnotemark[7]} 
\\ \hline  
  \multicolumn{1}{c}{} & \multicolumn{7}{c}{$s$} \\ \hline
 $I$ & 0.9037 & 0.9037 & 0.9037 & 0.9037 & 0.9037 & 0.9026 & 0.7817  \\
  $\mu_0$ & 0.1395 & 0.2139  & 0.1099 & 0.0223 & 0.0599  & 0.0226 & -0.1027 \\
  $\mu_1$ & -0.0655 & -0.0370 & -0.1388 & -0.4537 & -0.5390 & -2.4879 & -4.3920  \\
  $\mu_2$ &  &  &  &  & -12.6664 & -25.7710  \\
  Max. AE & 0.0002 & 0.0006 & 0.001 & 0.006 & 0.007 &  0.03 & 0.06 \\ \hline
  \multicolumn{1}{c}{} & \multicolumn{7}{c}{$p$} \\ \hline  
 $\mu_0$ & -0.0122 & 0.0164 & -0.0002 & -0.0168 & -0.0355 & -0.1039 & -0.1579  \\
  $\mu_1$ & -0.0227 & 0.0289 & -0.2077 & -0.2721 & -0.2201  & -6.1973 & -2.0656 \\
  $\mu_2$ &  &  &  &   & -107.6068 & -13.2488 \\
  $\mu_3$ &  &  &  &  & -521.8636 &  \\
  Max. AE & 0.0001 & 0.0009 & 0.005 & 0.002 & 0.008  & 0.007  & 0.02
\end{tabular}
\end{ruledtabular}
\footnotetext[1]{Non-relativistic calculations from Ref.~\cite{D96}}
\footnotetext[2]{The differences between the KS eigenvalues obtained with the exact potential 
of Ref.~\cite{UG94}.}
\footnotetext[3]{ALDA calculation including all bound and unbound states from Ref.~\cite{GKSG98}}
\footnotetext[4]{ALDA calculation including 34 unbound states from Ref.~\cite{PGB00}}
\footnotetext[5]{Hybrid calculation including 34 unbound states from Ref.~\cite{BPG00}}
\footnotetext[6]{ALDA calculation with WY ground-state potential from Ref.~\cite{WCY05}}
\footnotetext[7]{Asymptotically corrected ALDA results from Ref.~\cite{WCY05}}
\end{table*}
In this section we give the values of the best fit for He as defined
in Eq.~\ref{eq:fit} and as described thereafter. The original
excitation energies from which the quantum defects and corresponding
coefficients are calculated are obtained from the same sources as the
Be data described above. In Table~\ref{tab:HetddftabcS}, we give
results for He. The sizes of the quantum defects are much smaller than
in Be, and the fractional change between KS and experiment is
concomitantly larger. In the case of the $p$ quantum defects, they can
have large opposite signs. Application of ALDA to the exact
ground-state KS potential again works well.

In the case of He, the truncated-hybrid result improves upon the
truncated-ALDA. As we mentioned before this is because the hybrid was
developed for 2-electron systems.  AC-LDA and WY coefficients of He really
stand out as behaving much differently from the other cases and giving
large errors that cannot be removed by adding more coefficients. The
AC-LDA and WY coefficients of Be did not have this problem.

\subsection{\label{sec:triplets}Triplets}
\begingroup\squeezetable
\begin{table}
\caption{\label{tab:tddftabcT} The $\mu_i$ from different triplet TDDFT values 
for He and Be.}
\begin{ruledtabular}
\begin{tabular}{ccdddd}
\multicolumn{3}{c}{} & \multicolumn{1}{c}{} & \multicolumn{1}{c}{Truncated-} 
& \multicolumn{1}{c}{Truncated-} \\
  \multicolumn{2}{c}{} & \multicolumn{1}{c}{Ref.\footnotemark[1]} & \multicolumn{1}{c}{ALDA\footnotemark[2]} & 
\multicolumn{1}{c}{ALDA\footnotemark[3]} & \multicolumn{1}{c}{Hybrid\footnotemark[4]} 
 \\\hline  
  \multicolumn{2}{c}{} & \multicolumn{4}{c}{$s$} \\\hline
  He & $\mu_0$ & 0.2965  & 0.2719 & 0.2570 &  0.3171 \\
 & $\mu_1$ & -0.0811 &  -0.0892 & -0.1281 &   \\
 & Max. AE & 0.0001 & 0.004 & 0.003 & 0.007 \\\hline
  Be & $\mu_0$ & 0.7742 & 0.7706 & 0.7641 & 0.7683  \\
 & $\mu_1$ & 0.8870 & -0.3912  & -0.4015 & -0.2607 \\
& Max. AE & 0.001 &  0.0003 & 0.002 & 0.002  \\\hline
  \multicolumn{2}{c}{} & \multicolumn{4}{c}{$p$} \\\hline  
  He & $\mu_0$ & 0.0684 & 0.0734 & 0.0657 & 0.05332   \\
 & $\mu_1$ & 0.0457 & &  &  0.1103   \\
 & Max. AE & 0.0002 & 0.003 & 0.005 & 0.002  \\\hline
  Be & $\mu_0$ & 0.3620  &  0.4152 & 0.4101 & 0.37852 \\
 & $\mu_1$ & -0.4492 & 0.3553 & -0.7166 & -0.6404   \\
 & $\mu_2$ & 1.5829 & &  & 1.5225    \\
 & Max. AE & 0.0003   & 0.004 & 0.006 & 0.007    
\end{tabular}
\end{ruledtabular}
\footnotetext[1]{He non-relativistic calculations from Ref.~\cite{D96} and 
Be and Ne experimental values from NIST~\cite{NIST}}
\footnotetext[2]{ALDA calculation including all bound and unbound states from Ref.~\cite{GKSG98}}
\footnotetext[3]{ALDA calculation including 34 unbound states from Ref.~\cite{PGB00}}
\footnotetext[4]{Hybrid calculation including 34 unbound states for He and 
38 unbound states for Be from Ref.~\cite{BPG00}}
\end{table}
\endgroup 
In practice, spin decomposed TDDFT is commonly used, and
allows prediction of singlet$\rightarrow$triplet transitions.  In
this section we discuss the coefficients of the quantum defect
expansion obtained from triplet excitation energies. In
Table~\ref{tab:tddftabcT} we show the fit coefficients for He and Be.
For the WY and AC-LDA methods there is no triplet data available.  For
the triplet case, the $\mu_0$'s are very well reproduced in all cases,
also the other coefficients are often close. Overall the data can in
most cases be reproduced by only two or three coefficients.

\section{\label{sec:conclusions}Conclusions}
In the first part of this paper we have shown that the quantum defect
is a valuable quantity when reporting Rydberg states. A plot of the
quantum defect can be more insightful than a list of excitation
energies. The quantum defect can also be fitted to an expansion around
$E=0$ and a finite number of expansion coefficients can fully describe
the Rydberg series. We calculated the quantum defects for He, Be, and
Ne with accurate xc-potentials and showed that the KS quantum defect
lies between the interacting singlet and triplet values.

We also studied approximate ground-state KS potentials, namely, the LB94 and OEP
potentials. We saw that the OEP results are very close to the exact KS
values. The LB94 underestimates the quantum defect in all cases.
Again the data can be fully described by a few coefficients.

In the final part of the paper we calculated the quantum defects from
available TDDFT excitation energies from the literature. We see that
the quantum defect really amplifies the error in these cases. The
TDDFT data can be described by only a few coefficients.

Overall we see that while having an asymptotically correct potential
guarantees the existence of a Rydberg series, it is not necessarily a
good one. This is especially the case for AC-LDA TDDFT and LB94 ground
state values, and to lesser extent even for the WY TDDFT values.

\begin{acknowledgments}
We thank Qin Wu for providing the ionization energies corresponding
to their AC-LDA and WY calculations. We also thank Robert van Leeuwen 
for useful discussions.
This work was supported by NSF Grant No. CHE-0355405.
\end{acknowledgments}


\bibliography{myrefers}

\end{document}